# Review of OCS gas-phase reactions in dark cloud chemical models

*Jean-Christophe Loison[1]\*, Philippe Halvick[1], Astrid Bergeat[1], Kevin M. Hickson[1] and Valentine Wakelam[2,3]*

\*Corresponding author: jc.loison@ism.u-bordeaux1.fr

[1] *Univ. Bordeaux, ISM, UMR 5255, F-33400 Talence, France*
[2] *CNRS, ISM, UMR 5255, F-33400 Talence, France*
[3] *Univ. Bordeaux, LAB, UMR 5804, F-33270, Floirac, France.*
[4] *CNRS, LAB, UMR 5804, F-33270, Floirac, France*

The association reaction $S + CO \rightarrow OCS + h\nu$ has been identified as being particularly important for the prediction of gas-phase OCS abundances by chemical models of dark clouds. We performed detailed ab-initio calculations for this process in addition to undertaking an extensive review of the neutral-neutral reactions involving this species which might be important in such environments. The rate constant for this association reaction was estimated to be several orders of magnitude smaller than the one present in current astrochemical databases. The new rate for this reaction and the introduction of other processes, notably $OH + CS \rightarrow OCS + H$ and $C + OCS \rightarrow CO + CS$, dramatically changes the OCS gas-phase abundance predicted by chemical models for dark clouds. The disagreement with observations in TMC-1 (CP) and L134N (N), suggests that OCS may be formed on grain surfaces as is the case for methanol. The observation of solid OCS on interstellar ices supports this hypothesis.

## 1 INTRODUCTION

Chemical models of astrochemical environments can be very complex. In particular, hundreds of species involved in thousands of gas-phase reactions need to be included in order to follow the evolution of molecular complexity in the interstellar medium (ISM). Most of the rate constants for these reactions have not been studied in the laboratory over the pertinent temperature range. Sulphur chemistry is particularly poorly understood, for various reasons. First, the reactions of sulphur atoms present in astrochemical networks are mainly reactions with radicals which are delicate to study experimentally, particularly at low temperature.

Moreover, the S-bearing radicals and molecules are not so important in combustion or in atmospheric chemistry and so reactions involving sulphur radicals and molecules have not been as widely investigated as their oxygen bearing counterparts. Despite these uncertainties, S-bearing molecules are widely used to probe molecular density in dark clouds and star forming regions and are used to predict their stage of evolution (Pratap *et al.* 1997, Charnley 1997, Hatchell *et al.* 1998, van der Tak *et al.* 2003, Wakelam *et al.* 2004, Wakelam *et al.* 2011). With the aim of improving our knowledge about dark cloud chemistry, sensitivity analyses have been carried out to identify the "key" gas-phase reactions for dense clouds (Wakelam *et al.* 2010). Among these, the association reaction S + CO → OCS + hν seemed particularly important for OCS formation. The importance of this reaction was also underlined for low-mass protostellar envelopes (Wakelam et al. 2004). Given its potential importance, we calculated the rate constant for this exothermic but spin-forbidden radiative association, and found that the value commonly used in astrochemical databases was overestimated by several orders of magnitude. This prompted us to investigate the other neutral-neutral reactions that could produce or destroy OCS in an attempt to obtain a clearer picture of OCS chemistry in dark clouds.

For this study we have to evaluate rate constants and branching ratios at 10K for various reactions. For rate constant evaluation at 10 K the presence of a barrier is critical. When no information was available, the presence, and the value of barriers to the entrance valley were calculated at the M06-2X/cc-pVTZ level using the Gaussian09 software package except for the S + CO reaction for which calculations were performed at a higher level as defined in section 2.1. When the reaction is found, experimentally or theoretically, to have a significant barrier on the path of minimum potential energy leading from reactants to products the rate constant is set to zero at 10 K. When no barriers are present, the value of the rate constant is estimated using long-range forces, mainly through dispersion interactions and neglecting temperature dependency, between the reactants (Stoecklin & Clary 1992, Georgievskii & Klippenstein 2005) taking into account the electronic degeneracy. The electronic degeneracy factors are calculated by applying the spin and orbital correlation rules to the potential energy surfaces that correlate the separated reactants with the separated products. To evaluate branching ratio, we performed statistical calculations of the microcanonical rate constants of the various steps of the mechanisms (Bergeat *et al.* 2009). Quantum chemical calculations using the Gaussian09 software package were performed at the M06-2X/cc-pVTZ level to provide all relevant geometries, frequencies and rotational constants used for kinetic calculations.

This paper is organized as follows. In the following sections (sections 2 and 3) we review the different formation and destruction reactions for OCS, and the impact of the new reactions/rate constants for dark cloud modeling is studied in section 4. Our conclusions are given in Section 5.

## 2 OCS PRODUCTION

### 2.1 $S(^3P) + CO(^1\Sigma^+) \rightarrow OCS(^1\Sigma^+) + h\nu$

The $S(^3P) + CO(^1\Sigma^+) \rightarrow OCS(^1\Sigma^+) + h\nu$ reaction is currently thought to be the major source of OCS in the ISM. Nevertheless, it is spin forbidden and must pass through an excited $OCS(^3A')$ intermediate followed by a spin forbidden electronic transition: $OCS(^3A') \rightarrow OCS(^1\Sigma^+) + h\nu(UV)$, or a spin-orbit induced $OCS(^3A') \rightarrow OCS(^1\Sigma^+)^{**}$ crossing followed by the $OCS(^1\Sigma^+)^{**} \rightarrow OCS(^1\Sigma^+) + h\nu(IR)$ transition to occur. There have been previous theoretical studies of this reaction. Some by the Sayos group at a moderate computational level (MP2/6-311(2d) (Hijazo et al. 1994), MNDO/CI level (Sayos et al. 1990) and MP4 level (Gonzalez et al. 1996)), and one more recently at much higher level (CCSD(T)/aug-cc-VQZ//CCSD(T)/aug-cc-VTZ) (Adriaens et al. 2010). These studies are in qualitatively good agreement even if there are some differences for the height of the activation barrier, between 6.7 and 18.7 kJ/mol, the 18.7 kJ/mol corresponding to the latest calculation at the highest level. If the interaction of $S(^3P)$ with CO gives rise only to repulsive potential energy surfaces (PES), (Sayos et al. 1990, Gonzalez et al. 1996, Adriaens et al. 2010) thus making the direct $S(^3P) + CO(^1\Sigma^+) \rightarrow OCS(^3A') \rightarrow OCS(^1\Sigma^+) + h\nu(UV)$ impossible at low temperature, the interaction of $S(^1D)$ with CO yields at least one attractive PES which correlates with the singlet ground state of OCS. As a consequence, a triplet-singlet crossing occurs necessarily on the path from the triplet ground state of the reactants down to the singlet ground state of OCS. If spin-orbit induced $OCS(^3A') \rightarrow OCS(^1\Sigma^+)^{**}$ crossing occurs, this crossing is the bottleneck of the association reaction $S(^3P) + CO \rightarrow OCS$, and it is necessary to determine the lowest energy point of the crossing seam between the triplet manifold and the singlet PES of the OCS ground state to estimate the crossing rate constant at low temperature.

As the various wavefunctions describing the $S(^1D) + CO$ system are notably multiconfigurational, we have selected the complete active space self-consistent field (CASSCF) method. (Werner & Knowles 1985) The correlation energy has been calculated with the internally contracted multireference configuration interaction method, along with the

Pople correction (MRCI+Q) for size-consistency (Werner & Knowles 1988). Two different active spaces have been used. For the CASSCF calculations followed by MRCI+Q calculation, we used a small active space collecting only the molecular orbitals which correlate the $2p$ atomic orbitals of carbon and oxygen, and the $3p$ ones of sulfur. For the CASSCF calculations with no subsequent MRCI+Q calculation, a full valence active space was used, with equal weight for the various states with the same multiplicity (singlet or triplet). For all calculations, we used the cc-pVQZ basis set and the MOLPRO 2009 program package. The Pople correction was chosen as it reproduced well the transition energy $^3P$-$^1D$ of the sulfur: 107 kJ/mol to be compared to experimental transition energy of 108 kJ/mol (after removal of the spin-orbit splitting from the experimental data (Ralchenko *et al.* 2006)). A comparison between available literature data and our MRCI+Q calculations for the equilibrium bond lengths and dissociation energies of the ground state of OCS is presented in Table 1. Except for the O($^1D$) + CS dissociation energy, a good quantitative agreement is observed.

For the singlet surfaces, three PESs arise from the reactants S($^1D$) + CO in colinear approach ($C_{\infty v}$ symmetry) with $^1\Sigma^+$, $^1\Pi$ and $^1\Delta$ symmetry, the $^1\Sigma^+$ PES being attractive, while the $^1\Pi$ and $^1\Delta$ PESs are strongly repulsive as shown by CASSCF calculations. In a non-colinear approach, there are five PES, among which three have the $^1A'$ symmetry and two the $^1A''$ symmetry, the lowest energy surface being a $^1A'$ PES. Three angular regions have been identified. If the S atom approaches CO toward the C end, then the interaction potential is attractive and leads to the global minimum of the OCS singlet ground state PES. If the approach is perpendicular to the CO internuclear axis, then the interaction potential is strongly repulsive. Finally, if S approaches CO toward the O end, then the interaction potential is slightly attractive and leads to a weakly bound linear COS structure. The corresponding potential well is 3.5 kJ/mol deep with the geometry $r_{CO}$=1.131 Å and $r_{SO}$=2.958 Å given by MRCI+Q calculations. The global shape of this PES is in agreement with a previous investigation (Murrell & Guo 1987) and also with the PESs of the ground states of both $CO_2$ and $CS_2$ (Zúñiga *et al.* 1999). Only the approach of the S atom toward the C end of CO for the lowest PES of symmetry $^1A'$ needs to be considered as calculations have shown that the four other singlet PESs are strongly repulsive for any angle of approach of S toward CO.

For the triplet surfaces, two PESs arise from the reactants S($^3P$) + CO in colinear approach ($C_{\infty v}$ symmetry) with $^3\Sigma^-$ and $^3\Pi$ symmetry, both strongly repulsive, the $^3\Sigma^-$ PES being the most repulsive one. In a non-colinear approach there are three PESs, one of $^3A'$ symmetry and the two other of $^3A''$ symmetry. As we are searching for the lowest energy path, we need to consider only the $^3A'$ PES and the lowest $^3A''$ PES. CASSCF Calculations show that the interaction potential is less repulsive for non-colinear approach for both PESs than for the colinear approach. Potential wells have been found on these two PESs (Hijazo et al. 1994, Gonzalez et al. 1996) with the O-C-S angle ($\theta_{OCS}$) close to 124°, but with energies of the equilibrium structures significantly higher than the $S(^3P)$ + CO dissociation limit.

It is only when the $S(^1D)$ atom approaches CO toward the C end that the energy of the $^1A'$ PES decreases enough to cross the repulsive $^3A'$ and $^3A''$ PESs at the lowest possible energy. When $\theta_{OCS}$ varies from 180° to 120°, the $^1A'$ interaction potential between $S(^1D)$ and CO becomes less attractive and finally becomes repulsive, while in contrast the repulsive interaction potential between $S(^3P)$ and CO is weakening. We find that the lowest energy crossing singlet/triplet point is for $\theta_{OCS}$ = 131.9° with a crossing point 30 kJ/mol above the triplet reactants $S(^3P)$ + CO. Table 2 reports the geometry and MRCI+Q energy of the lowest energy point of the singlet/triplet crossing seams between the $^1A'$ PES and the $^3A'$ or $^3A''$ PES. Considering the high level method used for these theoretical calculations, the good agreement between the calculated and experimental OCS dissociation energies (Table 1) as well as the good agreement with previous calculations (Hijazo et al. 1994, Sayos et al. 1990, Gonzalez et al. 1996, Adriaens et al. 2010), our new ab-initio calculations show that the $S(^3P) + CO(^1\Sigma^+)$ → $OCS(^3A') + h\nu$ and the $S(^3P) + CO(^1\Sigma^+)$ → $OCS(^3A')$ → $OCS(^1\Sigma^+)^{**}$ → $OCS(^1\Sigma^+)$ + h$\nu$(IR) reactions are in fact negligible at 10 K and cannot be a source of OCS in the ISM.

**2.2 The HOCS system:**

Among the neutral – neutral reactions producing the OCS molecule, reactions involving the H-O-C-S system play a substantial role. The schematic energy diagram of this system is presented Fig 1 using known thermochemistry data (Baulch et al. 2005) or ab-initio calculations ((Rice et al. 1993, Rice et al. 1994) and this present study (M06-2X/aug-cc-pVTZ level)). We review the main reactions creating OCS except for the SH + CO → H + OCS reaction which is endothermic.

**2.2.1 CH + SO**

Due to the very high reactivity of the CH radical and by comparison with the CH + $O_2$ reaction, there is little doubt that the CH + SO reaction will present no barrier and should proceed through CH insertion into the SO bond (Bocherel *et al.* 1996, Huang *et al.* 2002, Bergeat *et al.* 2001) leading to the OC(H)S intermediate. There are five spin allowed exit channels for the CH($^2\Pi$) + SO($^3\Sigma^-$) reaction: CO($^1\Sigma^+$) + SH($^2\Pi$) (-569 kJ/mol), H($^2$S) + OCS($^1\Sigma^+$) (-521kJ/mol), OH($^2\Pi$) + CS($^1\Sigma^+$) (-284 kJ/mol), S($^3$P) + HCO($^2$A') (-275 kJ/mol) and O($^3$P) + HCS($^2$A') (-47 kJ/mol). RRKM calculations show than the O + HCS reaction leads to almost equal H + OCS (55%) and SH + CO (45%) production, and small contributions from O + HCS, S + HCO and OH + CS (which are neglected) . Thecalculated capture rate constant has a high value due to the dipole-dipole interaction term . However the reactants correlate with two doublet and two quadruplet surfaces, but the products $^2$H + $^1$OCS correlate with one doublet surface and $^2$SH + $^1$CO with two doublet surfaces. If we consider that only two doublet surfaces are attractive then there is a 4/12 = 1/3 electronic degeneracy factor. The global rate constant is then estimated to be equal to $k = 2\times10^{-10}$ cm$^3$ molecule$^{-1}$ s$^{-1}$ at 10K, close to the CH + O$_2$ rate constant value at 10 K (Bocherel et al. 1996), with almost no temperature dependence, leading to $k_{CH+SO\rightarrow H+OCS}(T) = 1.1\times10^{-10}$ cm$^3$ molecule$^{-1}$ s$^{-1}$ and $k_{CH+SO\rightarrow SH+CO}(T) = 9\times10^{-11}$ cm$^3$ molecule$^{-1}$ s$^{-1}$ in the 10-300K range.

**2.2.2 O + HCS**

The O($^3$P) + HCS($^2$A') reaction should be very similar to the O + HCO reaction and the rate constant present in current networks is deduced from the well known O + HCO reaction. (Baulch *et al.* 2005) The experimental rate constant for the O + HCO rate constant is close to the calculated capture rate constant value considering only the doublet surface as being attractive and so a 2/6 = 1/3 electronic degeneracy factor applies. As the capture rate constant for O + HCS reaction is almost equal to the capture rate constant for O + HCO reaction, we can use the same value for the O + HCS and O + HCO reactions. There are four exothermic and spin allowed exit channels for the O($^3$P) + HCS($^2$A') reaction: CO($^1\Sigma^+$) + SH($^2\Pi$) (-517 kJ/mol), H($^2$S) + OCS($^1\Sigma^+$) (-521 kJ/mol), OH($^2\Pi$) + CS($^1\Sigma^+$) (-239 kJ/mol), S($^3$P) + HCO($^2$A') (-228 kJ/mol). We have performed ab-initio calculations at the M06-2X/aug-cc-pVTZ level, showing no barrier for the oxygen atom addition to the carbon atom on the doublet surface, leading to $^2$OC(H)S intermediate. Our ab-initio calculations also show that the $^2$OC(H)S intermediate leads to H + OCS and SH + CO with small exit barriers and well below the reactant energy, in very good agreement with the previous calculations. (Rice & Chabalowski 1994) RRKM calculations performed in this study show than the O + HCS reaction leads mainly to H + OCS and SH + CO with similar production and negligible S +

HCO and OH + CS production. Combining the capture rate constant value and these statistical calculations of the branching ratio lead us to estimate $k_{O+HCS \to H+OCS}(T) = 5 \times 10^{-11}$ cm$^3$ molecule$^{-1}$ s$^{-1}$ and $k_{O+HCS \to SH+CO}(T) = 5 \times 10^{-11}$ cm$^3$ molecule$^{-1}$ s$^{-1}$ in the 10-300K range.

As the HCS abundance will affect the OCS abundance through the O + HCS reaction, the H($^2$S) + HCS($^2$A') → H$_2$($^1\Sigma^+$) + CS($^1\Sigma^+$) reaction will play an important role in determining the OCS abundance. This reaction is not present in current astrochemical networks and should be added. There is one previous theoretical study of this spin allowed reaction (Yamada *et al.* 2002) showing no barrier for this reaction on the singlet surface. This result is in good agreement with the similar and well known H + HCO reaction ($k_{H+HCO \to H2+CO}$ = $1.5 \times 10^{-10}$ cm$^3$ molecule$^{-1}$ s$^{-1}$ in the 300-2500 K), and we propose the use of the same rate constant value for the H + HCS → H$_2$ + CS reaction down to 10 K so $k_{H+HCS \to H2+CS}(T) = 1.5 \times 10^{-10}$ cm$^3$ molecule$^{-1}$ s$^{-1}$ in the 10-1000 K.

### 2.2.3 S + HCO

The S + HCO reaction is similar to the O + HCO and O + HCS reactions. We performed ab-initio calculations at the M06-2X/aug-cc-pVTZ level showing no barrier for sulphur atom addition to the carbon atom leading to the same OC(H)S intermediate as for the O + HCS reaction . Evolution of the OC(H)S intermediate was calculated using RRKM theory and leads to 70% and 30% of H + OCS and SH + CO production respectively, the OH + CS product channel being low Despite the fact that the polarisability of the sulphur atom is larger than the O atom one, the capture rate constant is similar, equal to $1.2 \times 10^{-10}$ cm$^3$ molecule$^{-1}$ s$^{-1}$ at 300K leading us to estimate $k_{S+HCO \to H+OCS}(T) = 8 \times 10^{-11}$ cm$^3$ mol$^{-1}$ s$^{-1}$ and $k_{S+HCO \to SH+CO}(T) = 4 \times 10^{-11}$ cm$^3$ molecule$^{-1}$ s$^{-1}$.

### 2.2.4 OH + CS

The OH($^2\Pi$) + CS($^1\Sigma^+$) → OCS($^1\Sigma^+$) + H($^2$S) reaction is present in the UMIST database (ref) with a rate constant equal to k($T$)= $9.39 \times 10^{-14} \times (T/298)^{1.12} \times \exp(-800/T)$ cm$^3$ molecule$^{-1}$ s$^{-1}$ in the 26-300 K range. There is no experimental determination for this reaction and the rate constant should have been deduced from the OH + CO reaction. However, there are two reliable ab-initio calculations (Rice *et al.* 1993, Adriaens et al. 2010) indicating the absence a barrier for this reaction indicating that it may occur at low temperature. The reaction mechanism is different from the three previous reactions with the OH + CS reaction leading to an H-O-•C=S intermediate. As there is a direct way to form cis-HOCS, (Rice &

Chabalowski 1994) the role of the OH…CS van der Waals complex is very minor and the rate constant for this reaction should be close to the capture limit, dominated by dipole-dipole interactions but with a strong dispersion contribution, with a value close to $4\times10^{-10}$ cm$^3$ molecule$^{-1}$ s$^{-1}$ in the 10-300 K range (before taking into account the electronic adiabaticity). There are only two exit channels: H($^2$S) + OCS($^1\Sigma^+$) (-237 kJ/mol) and CO($^1\Sigma^+$) + SH($^2\Pi$) (-289 kJ/mol). For formation of SH + CO to occur, considerable rearrangement is required or direct isomerization from cis-HOCS to cis-HCSO through a high energy transition state (Rice & Chabalowski 1994). For the flux going through the OC(H)S intermediate, statistical calculations of the evolution of OC(H)S leads us to predict 70% and 30% of H + OCS and SH + CO production respectively which is an upper limit for SH + CO production. As there is a direct way to produce H + OCS and as the transition state from HOCS to OC(H)S is quite high in energy, the SH + CO branching ratio is lower than 30% and we estimate it equal to 20 %. The estimated rate constants are then estimated, with an electronic degeneracy factor of 1/2 (the reactants correlate with two doublet surfaces but the products with only one doublet surface) to be $k_{OH+CS \to H+OCS}(T) = 1.7\times10^{-10}$ cm$^3$ molecule$^{-1}$ s$^{-1}$ and $k_{OH+CS \to SH+CO}(T) = 3\times10^{-11}$ cm$^3$ molecule$^{-1}$ s$^{-1}$ in the 10-300K range.

As mentioned above, the SH + CO → H + OCS reaction will not occur at low temperature because it is endothermic by 52 kJ / mol.

**2.3 Related reactions not playing any role in OCS formation**

The C($^3$P) + SO$_2$($^1$A$_1$) reaction has been studied experimentally at room temperature with a high value for the rate constant ($7\times10^{-11}$ cm$^3$ molecule$^{-1}$ s$^{-1}$) (Dorthe *et al.* 1991, Deeyamulla & Husain 2006) indicating no barrier in the entrance channel. There are three very exothermic exit channels for this reaction: CO($^1\Sigma^+$) + SO($^3\Sigma^-$) (-526 kJ/mol), CO$_2$($^1\Sigma^+$) + S($^3$P) (-537 kJ/mol) and OCS($^1\Sigma^+$) + O($^3$P) (-309 kJ/mol). Due to the valence of the O and S atoms, the most stable intermediate is likely to be the O-C-S-O one, with probable evolution toward the CO + SO exit channel, with a minor contribution from OCS + O production due to the smaller exothermicity. Given that the formation of CO$_2$ + S from the O-C-S-O intermediate requires significant rearrangement, these products are unlikely. We recommend the products as given in the OSU or UMIST databases for the reaction C + SO$_2$, namely CO + SO.

The O($^3$P) + CCS($^3\Sigma^-$) reaction is likely to proceed without a barrier in the entrance channel through comparison with the O($^3$P) + CCO($^3\Sigma^-$) reaction. The reactants, O($^3$P) + CCS($^3\Sigma^-$), correlate with singlet, triplet and quintuplet surfaces. The CO + CS exit channel correlates with singlet surfaces, CO($^1\Sigma^+$) + CS($^1\Sigma^+$), but also with triplet surfaces considering the excited triplet CO(a$^3\Pi$) or CS(a$^3\Pi$). However, the OCS($^1\Sigma^+$) + C($^3$P) products correlate only with triplet states. In fact, in the experimental study of the O + CCO reaction, the authors observed excited product CO* molecules in five electronic states *(A$^1\Pi$, d$^3\Delta$, e$^3\Sigma^-$, I$^1\Sigma^-$, a$^3\Sigma^+$)*, and they also suggested that CO(a$^3\Pi$) production occurred, even if wasn't directly detected. (Bayes 1970) Considering the calculated enthalpy of CCS formation (+608 kJ/mol) (Kaiser *et al.* 2002), the OCS + C exit channel is exothermic by -278 kJ/mol. However the unpaired electrons in CCS are localized on the terminal carbon atom, so the O atom will probably attack at the terminal carbon atom to form an OCCS intermediate, which will likely dissociate directly to CO + CS, the most statistically favored channel. This is the case for the OCS + C reaction which gives CO + CS with a rate constant close to the capture rate limit indicating very little back dissociation of the COCS intermediate (Deeyamulla & Husain 2006, Dorthe et al. 1991). Consequently, we neglect the OCS + C exit channel in this study. A more precise study is necessary to estimate the contribution of the CCO + S exit channel if the triplet surfaces play an important role. This would mean that the OCCS intermediate could dissociate to yield three different sets of products; CO(a$^3\Pi$) + CS($^1\Sigma^+$) ($\Delta H_r$ = -190 kJ / mol), CO($^1\Sigma^+$) + CS(a$^3\Pi$) ($\Delta H_r$ = -408 kJ / mol) and S($^3$P) + CCO($^3\Sigma^-$) ($\Delta H_r$ = -197 kJ / mol).

The S + CCO reaction is similar to the O + CCS reaction. By comparison with the O + CCS reaction, the S + CCO will almost certainly react rapidly to produce only CO + CS and not OCS + C.

The O($^3$P) + CCCS($^1\Sigma^+$) reaction is also likely to occur without a barrier in the entrance channel. The reactants correlate with one triplet $^3$A' and two triplet $^3$A", leading to various exothermic spin allowed product channels over triplet surfaces: CO($^1\Sigma^+$) + CCS($^3\Sigma^-$) (-320 kJ/mol), S($^3$P) + CCCO($^1\Sigma^+$) (-214 kJ/mol), CS($^1\Sigma^+$) + CCO($^3\Sigma^-$) (-154 kJ/mol) and OCS($^1\Sigma^+$) + C$_2$($^3\Pi_u$) (-112 kJ/mol) (using the NIST-webbook database for the enthalpies of formation except for $\Delta H_f$(CCCS($^1\Sigma^+$)) = 568 kJ/mol (Petrie 1996), $\Delta H_f$(CCS($^3\Sigma^-$)) = 608 kJ/mol (Kaiser et al. 2002)). The OCS + C$_2$ formation is then very unlikely to happen as it corresponds to the least exothermic exit channel.

The SH($^2\Pi$) + NCO(X$^2\Pi$) reaction can eventually lead to the exothermic (-59 kJ/mol) and spin allowed OCS($^1\Sigma^+$) + NH($^3\Sigma^-$) exit channel. However, a theoretical study of the OH($^2\Pi$) + NCO(X$^2\Pi$) reaction (Campomanes *et al.* 2000) shows the presence of an energy barrier, on both the singlet and triplet surfaces. As result we infer the presence of similar barriers for the SH + NCO reaction which is therefore assumed not to occur at low temperature.

The O + H$_2$CS and S + H$_2$CO reactions are both likely to possess a barrier in the entrance channel by comparison with the well known O + H$_2$CO reaction. (Baulch *et al.* 1992)

The OH + HCS reaction should have no barrier in the entrance channel leading to H$_2$O + CS by comparison with the well known OH + HCO reaction. (Baulch et al. 1992)

The reactions

$$CS + O_2 \rightarrow OCS + O$$
$$CS_2 + OH \rightarrow OCS + SH$$
$$CS_2 + O \rightarrow OCS + S$$

have high to medium activation energies (Atkinson *et al.* 2004, Murrells *et al.* 1990) and are thus negligible at low temperature. They do not play any role in OCS production in the ISM.

## 3 OCS CONSUMPTION

In the osu or UMIST databases, the only efficient processes at low temperature involving OCS as a reactant are ion-molecule reactions or photodissociation processes. Among the ion molecule reactions, the most efficient should be the reaction of OCS with C$^+$, S$^+$, CH$_3^+$, H$_3^+$ and He$^+$ reaction. However some neutral-neutral reactions, not present in UMIST or osu databases, also affect the OCS abundance, the C + OCS $\rightarrow$ CO + CS being the most important one.

### 3.1 C + OCS

The reactants C($^3$P) + OCS($^1\Sigma^+$) correlate only with triplet surfaces and so produce CO or CS in their first triplet excited states: CO($^1\Sigma^+$) + CS(a$^3\Pi$) or/and CO(a$^3\Pi$) + CS($^1\Sigma^+$). This reaction has been measured at 298 K with values equal to $1.01\times10^{-10}$ cm$^3$ molecule$^{-1}$ s$^{-1}$ and $5.6\times10^{-10}$ cm$^3$ molecule$^{-1}$ s$^{-1}$ (Dorthe et al. 1991, Deeyamulla & Husain 2006). The high experimental values for this reaction strongly indicate the absence of a barrier even if the reactants correlate with products in excited states. Moreover, as the electronic population of the $J$=0 state of C($^3$P) increases when the temperature decreases, the rate constant will increase towards low temperature if not all the electronic states correlate with the products. So a rate constant equal to $k_{C+OCS \to CO+CS}(T) = 1\times10^{-10}$ cm$^3$ molecule$^{-1}$ s$^{-1}$ at 10 K should be considered as a minimum value for this reaction and we recommend the use of this value in the 10-300K range.

### 3.2 CH + OCS

The CH + OCS reaction can lead to several products:

CH + OCS → CO + HCS
        → CS + HCO
        → CO + CS + H

The rate constant for this reaction has been measured between 297 K and 667 K leading to $k(T) = 2.0\times10^{-10}\times\exp(190/T)$ cm$^3$ molecule$^{-1}$ s$^{-1}$. (Zabarnick *et al.* 1989) The reactants CH(X$^2\Pi$) + OCS($^1\Sigma^+$) correlate with CO($^1\Sigma^+$) + HCS($^2$A') / CS($^1\Sigma^+$) + HCO($^2$A'). However, due to the very large exothermicities of these two exit channels (-665 kJ/mol and -532 kJ/mol) associated with the low value of the H-CO and H-CS bonds, the main exit channel should be the formation of spin allowed CO($^1\Sigma^+$) + CS($^1\Sigma^+$) + H($^2$S) products. The high value of the rate constant, coupled with the negative temperature dependence, clearly shows that this reaction does not present any barrier. However we cannot extrapolate the experimental temperature dependent expression obtained between 297 K and 667 K down to very low temperature as this would yield unrealistically large rate constant values. As a result, we recommend rather to use the value at 300K, $k_{CH+OCS \to CO+CS+H}(T) = 4\times10^{-10}$ cm$^3$ molecule$^{-1}$ s$^{-1}$, in the entire 10-300K range.

### 3.3 CN, C$_2$H, C$_2$ + OCS

The CN(X$^2\Pi$) + OCS(X$^1\Sigma^+$) reaction has a high rate value at 296 K, (Park & Hershberger 1998) equal to $9.75\times10^{-11}$ cm$^3$ molecule$^{-1}$ s$^{-1}$, probably leading to NCS(X$^2\Pi$) +

CO($X^1\Sigma^+$) formation as the NCO($X^2\Pi$) + CS($X^1\Sigma^+$) exit channel is endothermic by +115 kJ/mol (Park & Hershberger 1998). Ab-initio calculations have been performed on this system showing the absence of a barrier in the entrance valley (Zhang *et al.* 2005). Even if the experimental rate constant given (Park & Hershberger 1998) seems somewhat low considering the strong dipole-dipole interaction, we recommend using this value, $k_{CN+OCS \rightarrow NCS+CO}(T) = 1 \times 10^{-10}$ cm$^3$ molecule$^{-1}$ s$^{-1}$ over the 10-300K range.

The C$_2$H($X^2\Sigma^+$) radical is isoelectronic with CN and has a similar (or greater) reactivity with atoms and molecules. Even if there is no data for the C$_2$H($X^2\Sigma^+$) + OCS($X^1\Sigma^+$) reaction, we have decided to adopt the same rate constant as the one obtained for the CN + OCS reaction. The only possible exit channel will be HCCS($X^2A''$) + CO($X^1\Sigma^+$) as the HCCO($X^2A''$) + CS($X^1\Sigma^+$) channel is endothermic by 27 kJ/mol (Baulch et al. 2005).

The C$_2$($X^1\Sigma^+$) + OCS($X^1\Sigma^+$) reaction may lead to the exothermic and spin allowed CCS($a^1\Delta$, $b^1\Sigma^+$) + CO($X^1\Sigma^+$) products (-163 / -136 kJ/mol). (Garand *et al.* 2008) However it is difficult to determine whether this reaction presents a barrier. The C$_2$($X^1\Sigma^+$) + CO$_2$($X^1\Sigma^+$) reaction seems to present a barrier, (Reisler *et al.* 1980) even if the CCO($a^1\Delta$, $b^1\Sigma^+$) + CO($X^1\Sigma^+$) exit channels are exothermic (-213 / -183 kJ/mol), but the C$_2$($a^3\Pi_u$) + CS$_2$($X^1\Sigma^+$) reaction does not present a barrier, (Huang *et al.* 2004) even if the $a^3\Pi_u$ state of C$_2$ is in general less reactive than the ground $X^1\Sigma^+$ state. The absence of barrier for the C$_2$($a^3\Pi_u$) + CS$_2$($X^1\Sigma^+$) reaction is likely related to the ability of a sulfur atom to be hypervalent, the intermediate species being the C-C-S-C-S one. So OCS may react with the same pathway leading to C-C-S-C-O without barrier.

**3.4 OCS reactions with barriers**

The exothermic and spin allowed reactions O($^3$P) + OCS($^1\Sigma^+$) → CO($^1\Sigma^+$) + SO($^3\Sigma^-$) and H($^2$S) + OCS($^1\Sigma^+$) → CO($^1\Sigma^+$) + SH($X^2\Pi$) have been studied experimentally and present significant barriers in the entrance valley. (Atkinson et al. 2004, Tsunashima *et al.* 1975, Lee *et al.* 1977) The N($^4$S) + OCS($^1\Sigma^+$) → CO($^1\Sigma^+$) + NS($X^2\Pi$) reaction is also exothermic (-178 kJ mol$^{-1}$) but it is spin-forbidden and has never been studied. However the reactivity of a nitrogen atom in its ground state, N($^4$S), is in general smaller than that of O($^3$P) and the production of excited state CO or NS which would be spin-allowed is endothermic. As a result, we can neglect these three reactions at low temperature. The CS($X^1\Sigma^+$) + OCS($X^1\Sigma^+$)

→ $CS_2(X^1\Sigma^+)$ + $CO(^1\Sigma^+)$ reaction is exothermic (-125 kJ/mol) and spin-allowed. Although $CS(X^1\Sigma^+)$ is seen to react fairly rapidly with ground state atomic oxygen (albeit over a small barrier), (Lilenfeld & Richardson 1977) it is nevertheless seen to be relatively unreactive towards stable molecules (Black *et al.* 1983). As a result, we suggest that this process will be unimportant at low temperature. The $NCO(X^2\Pi)$ + $OCS(X^1\Sigma^+)$ → $CO_2(X^1\Sigma^+)$ + $NCS(X^2\Pi)$ reaction is also spin-allowed and exothermic (-64 kJ/mol). Despite the fact that the $NCO(X^2\Pi)$ radical in general displays high reactivity towards both unstable (Gao & Macdonald 2003) and stable (Becker *et al.* 2000) radical species, earlier measurements show it is unreactive towards $CO_2$ with an upper limit for the rate constant at 298 K of $1 \times 10^{-15}$ $cm^3$ $molecule^{-1}$ $s^{-1}$. (Becker *et al.* 1997) For this reason, we predict that the reaction with OCS will not play a role at low temperature. The $SO(^3\Sigma^-)$ + $OCS(X^1\Sigma^+)$ reaction has been indirectly observed to be very slow at room temperature, (Herron & Huie 1980) and so it should play no role at low temperatures. All possible exit channels for the $HCO(^2A')$ / $HCS(^2A')$ + $OCS(X^1\Sigma^+)$ reactions are endothermic and therefore these reactions will not take place at low temperatures either.

**4 Implications for chemical models of dark clouds**

To estimate the impact of these new proposed reactions and rate constants on the predicted abundance of OCS in dark clouds, we have used the latest version of the Nahoon chemical model (publicly available here: http://kida.obs.u-bordeaux1.fr/models/Nahoon_public_apr2011.tar.gz , see also Wakelam et al. in preparation).

This model computes the evolution of chemical abundances in the gas only. Grains are only considered to form molecular hydrogen and to undergo neutralization with cations when they carry negative charges. The chemistry is described by the kida.uva.2011 chemical network (http://kida.obs.u-bordeaux1.fr/models), which contains 6088 reactions and 474 species. Note that the reaction C + OCS had already been introduced in this network. To emphasize the effect of the proposed modification, we have used a modified version of kida.uva.2011, named kida.uva.2011* in which we have removed this reaction. This network is the next generation of kinetic database for cold environments initially based on the osu database (see Wakelam et al. in preparation). To show the effect of the proposed changes ("New values" in Table 3), we have calculated OCS abundances with several different networks. Elemental abundances and initial conditions are the same as in Wakelam & Herbst (Wakelam & Herbst 2008)(EA3 from Table 1). Typical dark cloud conditions are used: temperature of 10 K, H density of $2\times10^4$ cm$^{-3}$, cosmic-ray ionization rate of $1.3\times10^{-17}$ s$^{-1}$ and visual extinction of 10.

The abundance of OCS computed using kida.uva.2011* as a function of time is shown in Fig. 2 (model a; solid line). We also show in this figure the effects of changing the rate constants for: b the OH + CS reaction alone ; c the S + CO reaction alone; d the C + OCS reaction alone; e the OH + CS, S + CO and C + OCS reactions ; f all reactions from Table 3. Note that the reactions CN + OCS and $C_2H$ + OCS have not been included because the products NCS and HCCS are not currently in astrochemical databases. The modifications of the gas-phase reactions for OCS have a dramatic impact on the OCS abundance for typical dark cloud ages (up to $4\times10^5$yr). At $10^5$ yr, the OCS gas-phase abundance is only $10^{-11}$ (compared to the proton density), i.e. two orders of magnitude smaller than the abundance of $10^{-9}$ observed in the dark cloud L134N (North Peak). (Ohishi *et al.* 1992)

Among the reactions that we have included or modified, only three have a significant impact: S + CO → OCS + hν, C + OCS → CO + CS and OH + CS → OCS + H. It is then important to quantify the sensitivity of the modeled abundance to these new rate constants. For the association reaction S + CO → OCS + hν, our new theoretical calculations show that this reaction is negligible at 10K and should be removed from the network. For the C + OCS → CO + CS and OH + CS → OCS + H reactions we have estimated from previous experimental results and theoretical calculations (see 2.2.4 and 3.1) that the rate constants are known to within a factor of 2 and 3 respectively. We have plotted on Fig. 3 the OCS abundance computed with the new network by varying the new rate constants for the C + OCS (left) and OH + CS (right) reactions within their uncertainty limits. The OCS abundance depends linearly on the C + OCS rate constant between these limits from $4 \times 10^3$ to $10^5$ yr., and has a dramatic effect on the OCS abundance. Only the newly included OCS production reaction OH + CS → OCS + H prevents its abundance from dropping after $10^6$ yr. The variation of the OH + CS rate constant leads to an observable effect at all times, although the effect is more pronounced at later times. At $10^6$ yr, the OCS abundance changes by a factor of two when the OH + CS rate constant is multiplied or divided by three. The modification of the chemical network according to Table 3 has no significant impact on the other molecules, including S-bearing species.

## 5 Conclusions

Based on the identification of key reactions for dense clouds and protostellar envelopes, the association reaction S + CO → OCS + hv appeared to be crucial for the formation of OCS, despite the low value of the rate constant ($1.6 \times 10^{-17} (300/T)^{-1.5}$ cm$^3$s$^{-1}$ (Prasad & Huntress 1980)). The rate constant at 10 K for this process has been reevaluated, based on the presence of a notable barrier in the entrance valley, showing that the value used in astrochemical models was too high by several orders of magnitude. Decreasing this rate constant has a strong impact on the amount of OCS produced in the gas-phase. Other reactions have been studied in the context of this work. New reactions, absent from current astrochemical databases, and corrections of existing rate constants have been proposed. Among these reactions, two are of particular importance to the OCS abundance, one for the formation OH + CS → OCS + H and one for the destruction C + OCS → CO + CS, the latter having the most dramatic effect. The OCS gas-phase abundance predicted by our chemical model for dark cloud conditions is now much lower than the abundance observed in the two clouds TMC-1 (CP) and L134N (N). Previous chemical models reproducing the observations but the gas-phase chemistry was incorrect. This new result strongly suggests that OCS is produced on grain surfaces and is released into the gas-phase by non thermal processes, as is the case with methanol (Garrod *et al.* 2007). This hypothesis is strengthened by the observation of solid OCS at the surface of interstellar grains with abundances of about $10^{-7}$ (Palumbo *et al.* 1997), i.e. larger than the one observed in the gas-phase (about $10^{-9}$) and also by recent theoretical work. (Adriaens et al. 2010)

All the reactions and rate constants discussed in this paper will be included in the online KInetic Database for Astrochemistry (KIDA, http://kida.obs.u-bordeaux1.fr/) with their associated datasheets and we encourage astrophysicists to include these updated values in their models.

Acknowledgements:
The authors thanks the following funding agencies for their partial support of this work: the French CNRS/INSU program PCMI, the Agence Nationale de la Recherche (ANR-JC08−311018: EMA:INC) and the Observatoire Aquitain des Sciences de l'Univers.

Table 1. Equilibrium bond lengths and dissociation energies (without ZPE) calculated with the MRCI+Q method. Experimental data are given in parenthesis.

| Species | $r_{CS}$ (Å) | $r_{CO}$ (Å) | Energy (kJ/mol) |
|---|---|---|---|
| OCS($X^1\Sigma^+$) | 1.565 (1.560[a]) | 1.159 (1.160[a]) | 0.0 |
| S($^3P$)+CO($X^1\Sigma^+$) | … | 1.131 (1.128[b]) | 297 (295[c]) |
| S($^1D$)+CO($X^1\Sigma^+$) | … | … | 404 (404[c]) |
| O($^1D$)+CS($X^1\Sigma^+$) | 1.541 (1.541[b]) | … | 858 (838[c]) |

[a] (Herzberg 1966)

[b] (Huber & Herzberg 1979)

[c] Calculated with the reaction enthalpies at 0 K (L. V. Gurvich *et al.* 1989, Cox J.D. *et al.* 1989, Ruscic B. *et al.* 2006, Eland & Berkowitz 1979) from which the ZPE (Huber & Herzberg 1979, Herzberg 1966) has been subtracted and with the $^3P$-$^1D$ transition energies (Ralchenko et al. 2006) of O and S from which the spin-orbit splitting has been removed.

Table 2. MRCI+Q energy and geometry of the lowest energy point of the singlet/triplet crossing seams of the S + CO energy potential surfaces, the energy zero being the S($^3$P) + CO(X$^1\Sigma^+$) energy level.

| Crossing | $r_{CS}$ (Å) | $r_{CO}$ (Å) | $\theta_{OCS}$ (degree) | Energy (kJ/mol) |
|---|---|---|---|---|
| $^1$A' / $^3$A' | 2.256 | 1.135 | 131.9 | 30 |
| $^1$A' / $^3$A" | 2.295 | 1.134 | 137.4 | 32 |

Table 3: Summary of neutral-neutral reaction review. The "New values" are the results of the evaluation of the rate constants performed in this work.

| Reaction | UMIST (udfa06) | | | | OSU2003 | | | | New values | | | |
|---|---|---|---|---|---|---|---|---|---|---|---|---|
| | α | β | γ | F | α | β | γ | F | α | β | γ | F |
| **OCS production:** | | | | | | | | | | | | |
| O + HCS → H + OCS | 5.00e-11 | 0 | 0 | 2 | 5.00e-11 | 0 | 0 | 2 | 5.00e-11 | 0 | 0 | 2 |
| OH + CS | 5.00e-11 | 0 | 0 | 2 | 5.00e-11 | 0 | 0 | 2 | 0. | 0 | 0 | |
| SH + CO | 0 | 0 | 0 | | 0 | 0 | 0 | 0 | 5.00e-11 | 0 | 0 | 2 |
| S + HCO → H + OCS | - | - | - | - | - | - | - | - | 8.00e-11 | 0 | 0 | 2 |
| SH + CO | - | - | - | - | - | - | - | - | 4.00e-11 | 0 | 0 | 2 |
| OH + CS → H + OCS | 9.39e-14 | 1.12 | 800 | 2 | - | - | - | | 1.70e-10 | 0 | 0 | 3 |
| CO + SH | - | - | - | | - | - | - | | 3.00e-11 | 0 | 0 | 3 |
| CH + SO → OCS + H | - | - | - | | - | - | - | | 1.10e-10 | 0 | 0 | 2 |
| CO + SH | - | - | - | | - | - | - | | 9.00e-11 | 0 | 0 | 2 |
| | | | | | | | | | | | | |
| S + CO → OCS | 1.6e-17 | -1.5 | 0 | 1.25 | 1.6e-17 | -1.5 | 0 | 10 | 0 | - | - | 1 |
| SH + CO → H + OCS | | | | | | | | | 0 | - | - | 1 |
| O + CCS → CO + CS | - | - | - | - | - | - | - | - | 1.00e-10 | 0 | 0 | 2 |
| C + OCS | - | - | - | - | - | - | - | - | 0 | - | - | - |
| S + CCO → CO + CS | - | - | - | - | - | - | - | - | 1.00e-10 | 0 | 0 | 2 |
| C + OCS | - | - | - | - | - | - | - | - | 0 | - | - | 1 |
| C + SO$_2$ → OCS + O | 0 | | | | 0 | | | | 0 | | | |
| CO + SO | 7.00e-11 | 0 | 0 | 2 | 7.00e-11 | 0 | 0 | | 7.00e-11 | 0 | 0 | |
| | | | | | | | | | | | | |
| **OCS consumption:** | | | | | | | | | | | | |
| C + OCS → CO + CS | - | - | - | | - | - | | - | 1e-10 | 0 | 0 | 2 |
| CH + OCS → CO + CS + H | - | - | - | | - | - | | - | 4e-10 | 0 | 0 | 2 |
| CN + OCS → CO + NCS | - | - | - | | - | - | | - | 1e-10 | 0 | 0 | 2 |
| C$_2$H + OCS → CO + HCCS | - | - | - | | - | - | | - | 1e-10 | 0 | 0 | 3 |
| C$_2$ + OCS → CO + CCS | - | - | - | | - | - | | - | 1e-10 | 0 | 0 | 3 |
| | | | | | | | | | | | | |
| **Related reactions :** | | | | | | | | | | | | |
| H + HCS → H$_2$ + CS | - | - | - | - | - | - | - | - | 1.50e-10 | 0 | 0 | 2 |

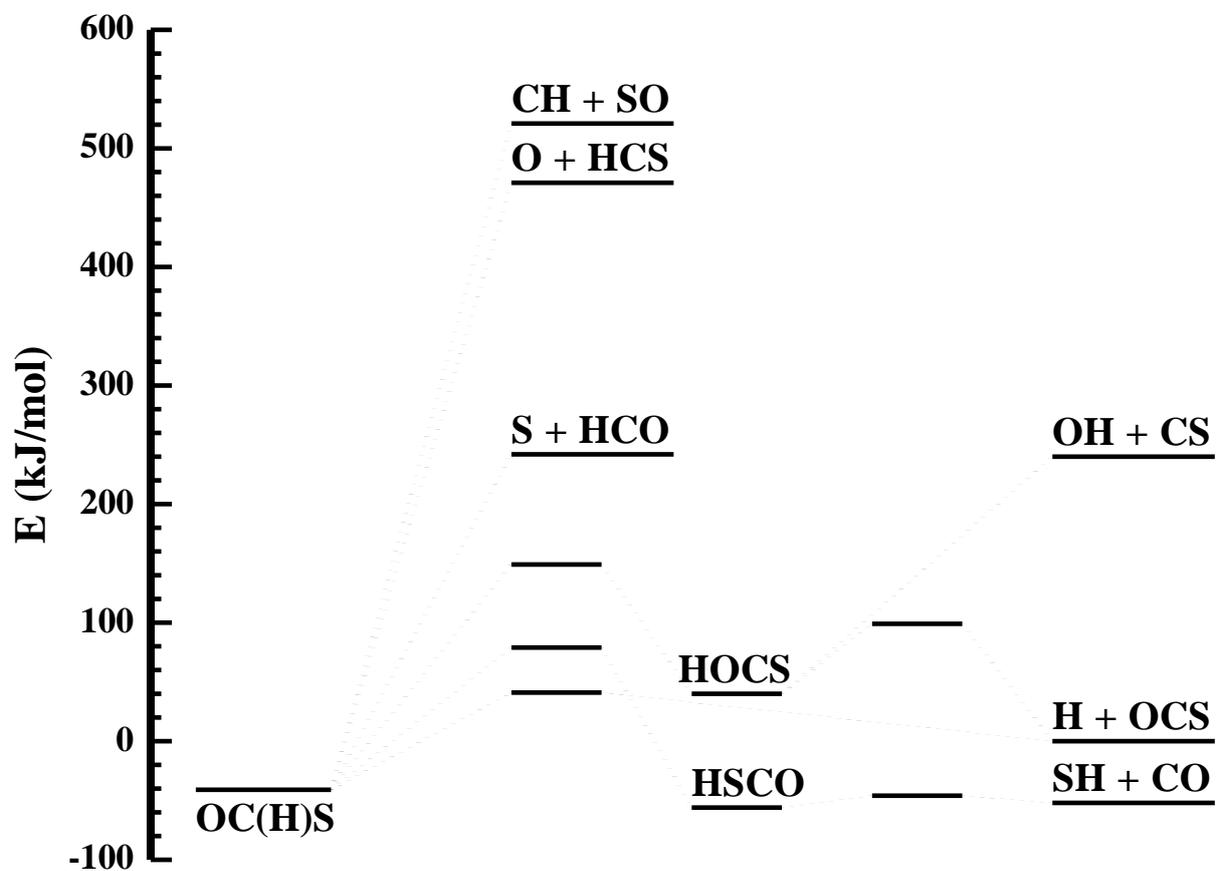

Fig 1. The schematic energy diagram (in kJ/mol) of the HOCS system on the doublet surface, the energy zero being that of the H + OCS energy.

Fig. 2: Abundance of OCS as a function of time for dark cloud conditions predicted by the model using different networks. Model a is the abundance computed with the kida.uva.2011* chemical network. For the other curves, reactions have been changed according to Table 3: b the OH + CS reaction alone ; c the S + CO reaction alone; d the C + OCS reaction alone; e the OH + CS, S + CO and C + OCS reactions ; f all reactions from Table 3.

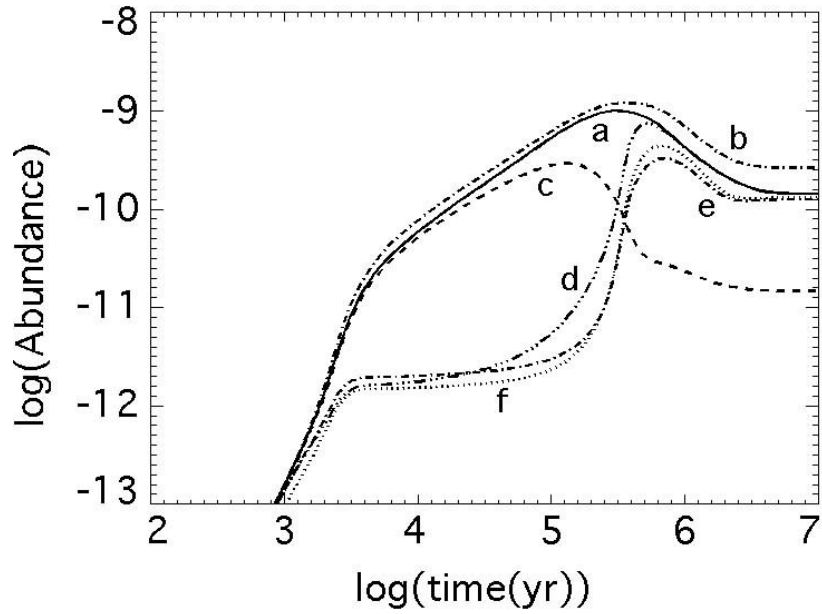

Fig. 3: Abundance of OCS computed as a function of time for dark cloud conditions. The solid line was obtained with the kida.uva.2011* chemical network (model a of Fig. 2) whereas the dashed line is for the updated network according to Table 3 (model f of Fig. 2). Dotted and dashed lines have been obtained with the updated network and the rate constant of the C + OCS (left panel) and OH + CS (right panel) reactions multiplied and divided by factors of two and three respectively.

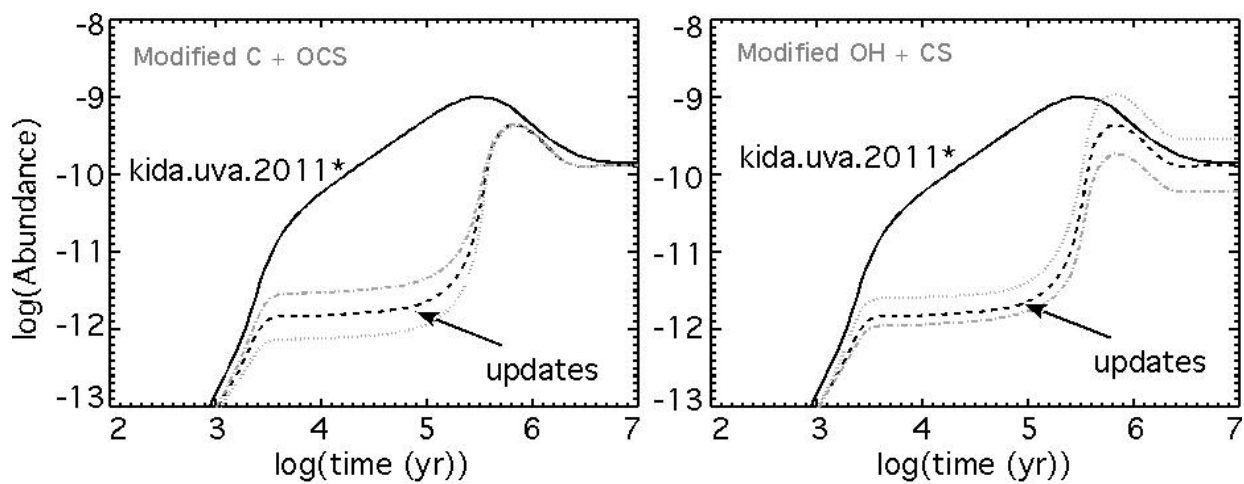